\documentclass[runningheads]{llncs}
\usepackage{moresize}
\usepackage{tugraz_defaults}
\usepackage{enumitem}
\usepgfplotslibrary{dateplot}
\usepgfplotslibrary{fillbetween}

\renewcommand{\paragraph}[1]{\vspace{0.0cm}\noindent\textbf{#1.}\ }
\renewcommand{\subparagraph}[1]{\vspace{0.0cm}\noindent\textbf{#1.}\ }

\newcommand{\parhead}[1]{\vspace{0.0cm}\noindent\textbf{#1.}\ }

\begin{document}

\date{}

\title{Practical Enclave Malware with Intel SGX}

\author{Michael Schwarz, Samuel Weiser, Daniel Gruss}
\institute{Graz University of Technology}

\maketitle

\pagestyle{empty}

\begin{abstract}
Modern CPU architectures offer strong isolation guarantees towards user applications in the form of enclaves. 
For instance, Intel's threat model for SGX assumes fully trusted enclaves, yet there is an ongoing debate on whether this threat model is realistic.
In particular, it is unclear to what extent enclave malware could harm a system. 
In this work, we practically demonstrate the first enclave malware which fully and stealthily impersonates its host application.
Together with poorly-deployed application isolation on personal computers, such malware can not only steal or encrypt documents for extortion, but also act on the user's behalf, \eg sending phishing emails or mounting denial-of-service attacks. 
Our SGX-ROP attack uses new TSX-based memory-disclosure primitive and a write-anything-anywhere primitive to construct a code-reuse attack from within an enclave which is then inadvertently executed by the host application.
With SGX-ROP, we bypass ASLR, stack canaries, and address sanitizer.
We demonstrate that instead of protecting users from harm, SGX currently poses a security threat, facilitating so-called super-malware with ready-to-hit exploits. 
With our results, we seek to demystify the enclave malware threat and lay solid ground for future research on and defense against enclave malware.
\end{abstract}

\section{Introduction}
\label{sec:intro} 

Software isolation is a long-standing challenge for the security of a computing system, especially if parts of the system are considered vulnerable, compromised, or even malicious~\cite{Intel_SGX2}.
Recent isolated-execution technology such as Intel Software Guard Extension~(SGX)~\cite{Intel_vol3} can shield software modules via hardware protected enclaves even from privileged kernel malware.
Thus, SGX has been advertised as key enabler of trusted cloud computing, where customers can solely rely on the CPU hardware for protecting their intellectual property and data against curious or malicious cloud providers~\cite{Schuster2015}.
Another use case for SGX is protecting copyrighted material from piracy~\cite{Bauman2016,Noubir2016} via DRM enclaves.
Also, SGX enclaves are explored for various further use cases, such as protecting ledgers~\cite{Ledger2017}, cryptocurrency wallets~\cite{Ledger2017}, password managers~\cite{Vrancken2017} and messengers~\cite{Signal2017}.
With the upcoming version SGXv2~\cite{Intel_vol3}, Intel opens their technology for the open-source community, allowing them to bypass Intel's strict enclave signing policy via their own key infrastructure.

However, there is a flip side to the bright future of isolated-execution technology painted by both the industry and community.
Any isolation technology might also be maliciously misused.
For instance, virtual machine extensions have been used to hide rootkits~\cite{King2006,Myers2007} and exploit CPU bugs~\cite{Weisse2018foreshadowNG}.
Researchers have warned that enclave malware likely causes problems for today's anti-virus~(AV) technology~\cite{Rutkowska2013,Davenport2014,Costan2016}.
The strong confidentiality and integrity guarantees of SGX fundamentally prohibit malware inspection and analysis, when running such malware within an enclave.
Moreover, there's a potential threat of next-generation ransomware~\cite{Marschalek2018Bluehat} which securely keeps encryption keys inside the enclave and, if implemented correctly, prevents ransomware recovery tools. Although there are few defenses proposed against potential enclave malware, such as analyzing enclaves before loading~\cite{Costan2016} or inspecting their I/O behavior~\cite{Davenport2014,Costan2016}, they seem too premature to be practical~\cite{Marschalek2018Bluehat}. 
Unfortunately, there exist no practical defenses against enclave malware, partly due to the lack of a proper understanding and evaluation of enclave malware.

\parhead{(Im-)Practicality of Enclave Malware}
Is enclave malware impractical anyway due to the strict enclave launch process~\cite{Intel_SGXSigning}, preventing suspicious enclave code from getting launch permission?
It is not, for at least four reasons:
First, adversaries would only distribute a benign-looking loader enclave, receiving and decrypting malicious payloads at runtime~\cite{Rutkowska2013,Marschalek2018Bluehat}. 
Second, Intel does not inspect and sign individual enclaves but rather white-lists signature keys to be used at the discretion of enclave developers for signing arbitrary enclaves~\cite{Intel_SGXSigning}.
In fact, we have a report from a student who independently of us found that it is easy to go through Intel's process to obtain such signing keys.
Third, the flexible launch control feature of SGXv2 allows bypassing Intel as intermediary in the enclave launch process~\cite{Intel_vol3}. 
Fourth, by infiltrating the development infrastructure of \emph{any} enclave vendor, be it via targeted attacks or nation state regulations, malware could be piggy-backed on their benign enclaves.
Hence, there are multiple ways to make enclave malware pass the launch process, with different levels of sophistication. 

\parhead{Impact of Enclave Malware}
Researchers have practically demonstrated enclave spyware stealing confidential information via side channels~\cite{Schwarz2017SGX}.
Apart from side-channel attacks, Costan~\etal~\cite{Costan2016} correctly argues that enclaves cannot do more harm to a system than an ordinary application process.
Yet, malware typically performs malicious actions from within an ordinary application process.
As an example, Marschalek~\cite{Marschalek2018Bluehat} demonstrated enclave malware which requires support of the host application to perform its malicious actions (\ie ransomware and shellcode).
No prior work has practically demonstrated enclave malware attacking a benign host application that does not collude with the enclave.
Hence, researchers believe that limitations in the SGX enclave execution mode severely restricts enclave malware in practice:
``Everyone's first reaction when hearing this, is `\textit{OMG bad guys will use it to create super malware!}'. But it shouldn't be that scary, because: Enclave programs are severely limited compared to normal programs: they cannot issue syscalls nor can they perform I/O operations directly.''~\cite{Aumasson2016}
Consequently, an enclave is believed to be limited by what its hosting application allows it to do:
``analyzing an application can tell you a lot about what an enclave can do to a system, mitigating the fear of a `\textit{protected malicious code running inside an enclave}'.''~\cite{Adamski2018}
At first glance, these statements seem reasonable, since syscalls are an essential ingredient for malware and enclaves can only issue syscalls through their host application. 
For example, Marschalek~\cite{Marschalek2018Bluehat} implemented enclave malware via a dedicated syscall proxy inside the host application to forward malicious enclave actions to the system.

In this work, we expand the research on enclave malware by presenting stronger enclave malware attacks.
As we show, enclave malware can overcome the SGX limitations. 
To that end, we develop a prototype enclave which actively attacks its benign host application in a stealthy way.
We devise novel techniques for enclaves probing their host application's memory via Intel TSX.
We find that enclave malware can effectively bypass any host application interface via code-reuse attacks, which we dub SGX-ROP.
Thus, the attacker can invoke arbitrary system calls in lieu of the host process and gain arbitrary code execution.
This shows that enclaves can escape their limited SGX execution environment and bypass any communication interface prescribed by their host.

We identify the core problem of research on enclave malware in a vagueness about the underlying threat model, which we seek to clarify in this work.
Intel's SGX threat model only considers \emph{fully trusted} enclaves running on an \emph{untrusted} host, which fits many scenarios like~\cite{Arnautov2016,Shinde2017Panoply,Tsai2017Graphene,Brenner2017SGX}. 
However, the asymmetry in this threat model ignores many other real-world scenarios, where enclaves might not be unconditionally trustworthy.
In particular, while the (third-party) enclave vendor might consider its own enclave trustworthy, the user or the application developer that use a third-party enclave both have all rights to not trust the enclave.s
To address this asymmetry, we introduce a new threat model which specifically considers untrusted enclaves. 
This allows to reason about attacks from within enclaves, such as \eg enclave malware, and to identify scenarios under which potential enclave malware becomes decisive.

\parhead{Contributions} We summarize our contributions as follows.
\begin{enumerate}[nolistsep]
\item We introduce a new threat model which considers malicious enclaves.
\item We discover novel and stealthy TSX memory probing primitives.
\item We present SGX-ROP, a practical technique for enclave malware to perform malicious operations, \eg on the system level, \textit{without} collaboration from the host application.
\end{enumerate}

The rest of the paper is organized as follows.
Section~\ref{sec:background} provides background.
Section~\ref{sec:threatmodel} describes our threat model.
Section~\ref{sec:attackoverview} overviews our attack.
Section~\ref{sec:tsx} shows how to locate gadgets and Section~\ref{sec:rop} shows how to use them.
Section~\ref{sec:evaluation} evaluates our attack.
Section~\ref{sec:discussion} provides a discussion.
Section~\ref{sec:conclusion} concludes.

\section{Background}
\label{sec:background} 
In this section, we overview address spaces, Intel SGX, TSX as well as control-flow attacks and trigger-based malware.

\parhead{Virtual Address Space}\label{sec:translation}
Modern operating systems rely on virtual memory as an abstraction layer to the actual physical memory.
Virtual memory forms the basis for process isolation between user applications and towards the kernel. 
Permissions are set on a granularity of pages, which are usually \SI{4}{\KB}.
Permissions include \emph{readable}, \emph{writable}, \emph{executable}, and \emph{user accessible}. 
On modern x86-64 CPUs, the usable virtual address space is $2^{48}$ bytes, divided into user space and kernel space, with $2^{47}$ bytes each. 
Within the user space of an application, the entire binary as well as all shared libraries used by the application are mapped. 

\parhead{Intel SGX}\label{sec:sgx}
Intel Software Guard Extension (SGX) is an instruction-set extension for protecting trusted code, introduced with the Skylake microarchitecture~\cite{Intel_vol3}. 
Applications are split into untrusted and trusted code, where the latter is executed within an enclave. 
The threat model of SGX assumes that the enclave environment, \ie operating system and all normal application code, might be compromised or malicious and cannot be trusted. 
Hence, the CPU guarantees that enclave memory cannot be accessed from any other part of the system, except for the code running inside the enclave. 
Enclaves can therefore safely run sensitive computations, even if the operating system is compromised by malware.
Still, memory-safety violations~\cite{Lee2017SGXROP}, race conditions~\cite{Weichbrodt2016}, or side-channel leakage within the enclave~\cite{Brasser2017sgx,Schwarz2017SGX} might lead to exploitation.

\begin{figure}[t]
 \centering
 \resizebox{0.6\hsize}{!}{
 \tikzstyle{process} = [rectangle, minimum width=2cm, minimum height=0.65cm, text centered, draw=black, fill=orange!30]
\tikzstyle{arrow} = [thick,->,>=stealth]
\begin{tikzpicture}[yscale=0.65]
\definecolor{msblue}{HTML}{0000FF}

\draw (0.25, 2.25) rectangle node [yshift=1.85cm] {\large Application} +(9,4.75);

\draw [fill=yellow!40] (5.75, 2.5) rectangle node [anchor=north,yshift=2.1cm,xshift=.7cm] {Trusted enclave} +(3, 4);
\draw [fill=red!40] (5.5, 4) rectangle node [rotate=90,color=black] {Call Gate} +(0.5,2.5);

\draw [fill=gray!20!white] (0.75, 2.5) rectangle node [anchor=north,yshift=2.1cm,xshift=-.7cm] {Untrusted} +(3, 4);
\node (create) at (2.25,6) [process] {\small Load Enclave};
\draw [arrow] (2.25,5.68) -- (2.25,4.82);
\node (call) at (2.25,4.5) [process] {\small Call Trusted Fnc.};
\draw [arrow] (2.25,4.18) -- (2.25,3.42);
\node (call) at (2.25,3.1) [process] {\small ... };

\draw [arrow] (3.55,4.65) -- (5.5,5.5);
\draw [arrow] (6,5.5) -- (6.5,5.5);
\draw (6.5,3.5) edge[out=180,in=0,arrow] (3.55,4.35);
 
\node (create) at (7.5,5.5) [process] {\small Trusted Code};
\node (create) at (7.5,3.5) [process] {\small Return};
\draw [arrow] (7.5,5.17) -- (7.5,3.82);



\end{tikzpicture} 
 }
 \caption{In the SGX model, applications consist of an untrusted host application and a trusted enclave. 
The hardware prevents any direct access to the enclave code or data. 
 The untrusted part uses the \texttt{EENTER} instruction to call enclave functions that are exposed by the enclave.
 }
 \label{fig:sgx}
\end{figure}
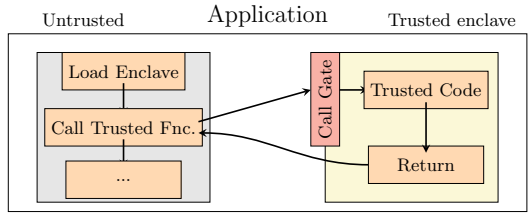

The integrity of an enclave is ensured in hardware by measuring the enclave loading process and comparing the result with the reference value specified by the enclave developer. 
Once loaded, the application can invoke enclaves only at defined entry points.
After the enclave finishes execution, the result of the computation, and the control flow, is handed back to the calling application. 
\Cref{fig:sgx} illustrates the process of invoking a trusted function inside an enclave. 

Enclave memory is mapped in the virtual address space of its host application. 
To allow data sharing between enclave and host application, the enclave is given full access to the entire address space of the host application.
This protection is not symmetric and gives raise to enclave malware.

\parhead{Hardware Transactional Memory}\label{sec:tsx-background}
Hardware transactional memory attempts to optimize synchronization primitives with hardware support.
The hardware provides instructions to create so-called \emph{transactions}. 
Any memory access performed within the transaction is not visible to the outside until the transaction is successfully completed, thus providing atomicity for memory accesses. 
However, if there is a conflict within the transaction, the transaction aborts and all changes done within the transaction are rolled back, \ie the previous state is restored. 
A conflict can be the concurrent modification of a data value by another thread, or an exception, \eg a segmentation fault. 

Intel TSX is an instruction-set extension that implements transactional memory by leveraging the CPU cache. 
TSX works on a cache line granularity, which is usually \SI{64}{B}. 
The CPU keeps track of a so-called read and write set. 
If a cache line is read or written inside the transaction, it is automatically added to the read or write set, respectively. 
Concurrent modifications to data in a read or write set from different threads cause a transaction to abort. The size of the read set, \ie all memory locations read inside a transaction, appears to be limited by the size of the L3 cache, and the size of the write set, \ie all memory locations modified inside a transaction, appears to be limited by the size of the L1 cache~\cite{Liu2014concurrent}. 

Transactional memory was described as a potential security feature, \eg by
Liu~\etal\cite{Liu2014concurrent} to detect rootkits, and by Guan~\etal\cite{Guan2015} to protect cryptographic keys when memory bus and DRAM are untrusted.
Kuvaiskii~\etal\cite{Kuvaiskii2016} showed that TSX can be leveraged to detect hardware faults and revert the system state in such a case. 
TSX also opens new attack vectors, \eg by 
Jang~\etal\cite{Jang2016} abusing the timing of suppressed exceptions to break KASLR.

\parhead{Control-Flow Attacks}\label{sec:cf-attacks}
Early control-flow attacks exploited a buffer overflow to inject arbitrary code into a program.
Modern CPUs allow to prevent these attacks by marking writable pages as non-executable~\cite{Intel_vol3}. Thus, an attacker has to resort to more complex so-called \emph{code-reuse attacks}.
Shacham~\etal\cite{Shacham2007} presented return-oriented programming (ROP), which abuses the stack pointer to control the instruction pointer. 
For this purpose, addresses of \emph{gadgets}, \ie very short code fragments ending with a \texttt{ret} instruction are injected into the stack. 
Whenever executing \texttt{ret}, the CPU pops the next gadget address from the stack and continues execution at this gadget. 
By stitching together enough gadgets, one can gain arbitrary code execution.

A mitigation against these attacks present in modern operating systems is to randomize the virtual address space. 
Address space layout randomization (ASLR)~\cite{PaxASLR} ensures that all regions of the binary are at random locations every time the binary is executed. 
Thus, addresses of gadgets are unpredictable, and an attacker cannot reliably reference gadgets anymore. 
Assuming no information leak and a large enough entropy (\eg on 64-bit systems), ROP attacks become infeasible, as addresses cannot even be guessed~\cite{Strackx2009breaking,Szekeres2013sok}. 
Furthermore, some techniques are deployed against such attacks, \eg stack canaries~\cite{Cowan1998,Pax2015rap}, shadow stacks~\cite{Chiueh2001}, stack-pivot defenses~\cite{Yan2016pivot}.

\parhead{Trigger-based Malware}\label{sec:trojans}
With increasing connectivity between computer systems in the past decades, malware evolved into a significant security threat.
Especially the rise of the internet facilitated the spreading of malware~\cite{Gupta2009empirical} over millions of affected systems.
There is a market for malware with various targets~\cite{Caulfield2017us,EFF2011NSAtrojan}.
In many cases, malware remains in an inactive state, until a specific time~\cite{Crandall2006temporal} or a remote command triggers activation~\cite{Sharif2008impeding,Andriesse2014instruction}.
This decorrelates attack from infection and enables synchronous attacks as well as targeted attacks (\eg activating the malware only on certain target systems).

The entry point for malware is often a vulnerability, whose exploitation (\eg via a control-flow attack) enables malicious operations on the target device.
While userspace malware then typically misuses lax privilege management of commodity operating systems to access user documents or impersonate user action, more sophisticated malware seeks to elevate privileges even further.

Exploits can rely on an undisclosed vulnerability~\cite{Egelman2013markets,Caulfield2017us}, making it very difficult to mitigate attacks. 
For certain actors, there is an interest in keeping such zero-day exploits undisclosed for a long time~\cite{Hall2017time}.
As a consequence, modern malware is obfuscated to remain stealthy~\cite{You2010malware}, \eg via code obfuscation~\cite{Sharif2008impeding}, or steganography~\cite{Andriesse2014instruction}. 
However, a thorough malware analysis may revert obfuscation~\cite{Jiang2007stealthy} and expose the underlying vulnerability.

\section{Threat Model}
\label{sec:threatmodel} 
In this section, we show limitations of the SGX threat model regarding malicious enclaves and present our threat model considering enclave malware. 

\subsection{Intel's SGX Threat Model}
In Intel's SGX threat model, the entire environment, including all non-enclave code is untrusted (\cf Section~\ref{sec:background}).
Such a threat model is primarily useful for cloud computing with sensitive data if a customer does not fully trust the cloud provider, and for protection of intellectual property (\eg DRM), secret data or even legacy applications inside containers~\cite{Arnautov2016,Shinde2017Panoply,Tsai2017Graphene,Brenner2017SGX}.  
With SGX, developers can use enclaves without the risk of exposing sensitive enclave data.

However, this model provides no means to protect other software, apart from enclaves themselves.
In particular, applications hosting enclaves are not protected against the enclaves they load.
Furthermore, enclaves cannot be inspected if they choose to hide their code, \eg using a generic loader. 
This asymmetry may foster enclave malware, as SGX can be abused as protection mechanism in addition to obfuscation and analysis evasion techniques.
One could argue that host applications themselves could be protected by means of additional enclaves.
However, this is not always feasible and even impossible for certain code.
Some reasons for keeping application code outside enclaves are the restricted feature set of enclaves (\eg no syscalls), expensive encrypted enclave-to-enclave communication, and an increased porting effort.
Hence, there are many practical scenarios, as we illustrate in which a host application might be threatened by an enclave, which are not covered by Intel's threat model.

\subsection{Our Threat Model Considering Enclave Malware}
\parhead{Victim}
In our threat model, we assume that a user operates a computing device which is the target of an attacker.
The user might be a private person, an employee or a system administrator in a company.
From the user's perspective, the system (including the operating system) is considered trusted and shall be protected against malware from third-party software.
The device has state-of-the-art anti-malware or anti-virus software installed for this purpose.
This applies to virtually all Windows systems today, as Windows 10 comes with integrated anti-virus software.
The user executes a benign application which depends on a potentially malicious (third-party) enclave.
The benign host application communicates with the enclave through a tight interface (\eg a single ECALL). 
This interface, if adhered to, would not allow the enclave to attack the application.
Furthermore, we assume that the host application is well-written and free of software vulnerabilities.
Also, the application incorporates some state-of-the-art defenses against runtime attacks such as ASLR and stack canaries.

\parhead{Attacker}
The attacker controls the enclave used by the host application, which we denote as the malicious enclave.
The attacker seeks to escape the enclave and gain arbitrary code execution with host privileges.
Also, the attacker wants to achieve plausible deniability until he chooses to trigger the actual exploitation, \ie the exploit should be undetectable until executed.
This decouples infection from exploitation and allows the attacker to mount large-scale synchronous attacks (\eg botnets, ransomware) or target individuals.
To that purpose, the attacker encloses malware in the enclave in a way that prevents inspection by any other party.
This can be done by receiving and decrypting a malicious payload inside the enclave at runtime via a generic loader~\cite{Rutkowska2013}, for example.

While the attacker can run most unprivileged instructions inside the enclave, SGX not only prevents enclaves from executing privileged instructions but also syscalls, among others~\cite{Intel_vol3, Intel_SGX2}. 
Moreover, enclaves can only execute their own code.
An attempt to execute code of its host application (\eg by using \texttt{jmp}, or \texttt{call}), results in a general protection fault, and, thus, termination of the enclave~\cite{Intel_SGX2}.
Thus, a successful attack must bypass these restrictions.
Finally, we assume that the attacker does not exploit potential hardware bugs in the SGX implementation (\eg CVE-2017-5691).  

\parhead{Scenarios}
We consider three scenarios, two with a criminal actor and one with a surveillance state agency~\cite{EFF2011NSAtrojan,Bundestrojaner2017Germany}.
In the first scenario, a criminal actor provides, \eg a computer game requiring to run a DRM enclave, or a messenger app requiring to run an enclave for security mechanisms~\cite{Signal2017}.
In the second, a criminal actor provides an enclave that provides an interesting feature, \eg a special decoder, and can be included as a third-party enclave.
In the last scenario, it may be an app the state endorses to use, \eg an app for legally binding digital signatures which are issued by an enclave, or legal interactions with authorities.
Also, in some countries, state agencies might be able to force enclave vendors to sign malicious enclaves on their behalf via appropriate legislation, \eg replacing equivalent benign enclaves.
In any case, the externally controlled enclave might perform unwanted actions such as espionage or hijacking of the user's computer.

\section{Attack Overview}\label{sec:attackoverview} 
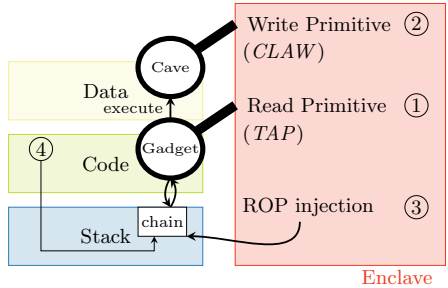
\begin{figure}[t]
\centering
\resizebox{0.5\hsize}{!}{
\begin{tikzpicture}[yscale=0.9]
\draw[fill=red!20,draw=red] (4,-2.5) rectangle +(3.25,4.5) node[yshift=-4cm,xshift=-0.75cm,below] {\textcolor{red}{Enclave}};

\draw[fill=green!20,draw=green] (0.5,-1.25) rectangle +(3,1) node[pos=.5] {Code};
\draw[fill=yellow!20,draw=yellow] (0.5,0) rectangle +(3,1) node[pos=.5] {Data};
\draw[fill=blue!20,draw=blue] (0.5,-2.5) rectangle +(3,1) node[pos=.5] {Stack};

\begin{scope}[xshift=3cm,yshift=-0.5cm]
\draw[line width=1.5mm] (0,0) -- (1,0.75) node[above,right] {Read Primitive};
\draw[line width=0.8mm,fill=white] (0,0) circle (0.5) node[pos=.5] {\scriptsize Gadget};
\node[anchor=west] at (3.5,0.75) {\scriptsize \circleds{1}};
\node[anchor=west] at (1,0.25) {(\emph{TAP})};
\end{scope}

\begin{scope}[xshift=3cm,yshift=0.9cm]
\draw[line width=1.5mm] (0,0) -- (1,0.75) node[above,right] {Write Primitive};
\draw[line width=0.8mm,fill=white] (0,0) circle (0.5) node[pos=.5] {\scriptsize Cave};
\node[anchor=west] at (3.5,0.75) {\scriptsize \circleds{2}};
\node[anchor=west] at (1,0.25) {(\emph{CLAW})};
\end{scope}

\node[anchor=west] at (4,-1.5) {ROP injection};
\node[anchor=west] at (6.5,-1.5) {\scriptsize \circleds{3}};

\draw (5,-1.75) edge[out=270,in=0,->,>=stealth,thick] node[below] {\scriptsize } (3.25, -2);
\draw (3, -1.5) edge[out=60,in=300,->,>=stealth,thick] node[left,yshift=0.25cm] {} (3, -1);
\draw (3, -1) edge[out=240,in=120,->,>=stealth,thick] node[left,yshift=0.25cm] {} (3, -1.5);

\draw[fill=white] (2.5, -2) rectangle +(0.75,0.5) node[pos=.5] {\scriptsize chain};

\draw (3, 0) edge[out=90,in=270,->,>=stealth,thick] node[left] {\scriptsize execute} (3, 0.4);

\node at (1,-0.5) {\scriptsize \circleds{4}};
\draw[->,>=stealth] (1,-0.7) |- (1,-2.25) -| (2.75, -2);

\end{tikzpicture}
}
\caption{Attack overview}
\label{fig:attack}
\vspace{-0.6cm}
\end{figure}
In this section, we outline how enclave malware can successfully attack a system using novel methods we discover. 
In particular, we show how enclave malware can evade all restrictions SGX poses on enclave execution.
This allows the enclave to run arbitrary code disguised as the host process, similar to process hollowing~\cite{Leitch2013process}, which is often used by malware. 
In fact, one can conceal existing user-space malware inside an SGX enclave, \eg ransomware.

\parhead{Restricted Enclave Environment}
In contrast to most traditional malware, a malicious enclave has to act blindly. 
SGX prevents enclaves from directly executing syscalls (\cf \Cref{sec:threatmodel}), an essential ingredient for user-space malware, and mandates the host application with this task. 
Also, the memory layout of the host application as well as its code base might be unknown to the enclave. 
Note that one enclave can be loaded by different host applications.
Enclaves only have knowledge of the ECALL/OCALL interface through which they communicate with the host.
Hence, the malicious enclave needs to assemble an attack without knowledge of the host application memory and without executing syscalls.

\parhead{Novel Fault-Resistant Primitives}
To overcome these restrictions, we leverage TSX and SGX to construct a \emph{fault-resistant read} primitive as well as a \emph{fault-resistant write-anything-anywhere} primitive.
While the read primitive helps in scanning host memory, the write primitive identifies writable memory which we denote as a cave. 
Those primitives are fault resistant in the sense that the enclave can safely probe both mapped and unmapped memory without triggering exception behavior that would abort the enclave.
By combining both primitives, the attacker can mount a code-reuse attack (\ie ROP) on the host application, which we call SGX-ROP. 

\parhead{SGX-ROP}The actual SGX-ROP attack is performed in four steps, as depicted in Figure~\ref{fig:attack}.
In step \circleds{1}, the malicious enclave uses the read primitive to scan the host application for usable ROP gadgets.
In step \circleds{2}, the enclave identifies writable memory caves via the write primitive and injects arbitrary malicious payload into those caves.
In step \circleds{3}, the enclave constructs a ROP chain from the gadgets identified in \circleds{1} and injects it into the application stack.
Then, the enclave returns execution to the host application and the attack waits to be activated. 
When the application hits the ROP chain on the stack, the actual exploitation starts (step \circleds{4}).
The ROP chain runs with host privileges and can issue arbitrary system calls.
While this is already sufficient for many attacks, we go one step further and execute arbitrary attack code in the host application by marking the cave (\cf step \circleds{2}) as executable and invoking the code stored in the cave.
After exploitation, the cave can eliminate any traces in the host application and continue normal program execution.

SGX-ROP works without the necessity of a software bug in the host application. 
The write primitive further allows to even bypass some anti-control-flow-diversion techniques (\cf \cref{sec:cf-attacks}) as any writable data can be modified.
This includes ASLR, stack canaries, and address sanitizer, which we all bypass with our attack (\cf \cref{sec:full-exploit}).

\section{Locating Code Gadgets}
\label{sec:tsx} 

In this section, we show how an enclave attacker can stealthily scan its host application for ROP gadgets.
The attacker does not need any a-priori knowledge about the host application memory layout.
We first discuss why existing memory scanning techniques are not applicable.
Next, we show how to use TSX to construct a novel fault-resistant memory disclosure primitive. 
Finally, we leverage this primitive to discover accessible code pages of the host application and subsequently leak the host application binary. 
This enables an attacker to search for ROP gadgets to construct the actual attack (\cf \Cref{sec:rop}). 

\subsection{Problem Statement}
The malicious enclave wants to scan host application memory to craft an SGX-ROP attack.
Luckily for the attacker, the SGX memory protection is asymmetric. 
That is, SGX prevents non-enclave code from accessing enclave memory, while an enclave can access the entire memory of the host application as they run in the same virtual address space.
Thus, the enclave naturally has a read primitive.
However, the enclave might not know anything about the host application's memory layout (\eg which pages are mapped, or their content), apart from the ECALL/OCALL interface.
The enclave cannot query the operating system for the host memory mapping (\eg via \texttt{/proc/pid/maps}), as system calls cannot be performed from within an enclave. 
The enclave could naively try to read arbitrary host memory.
However, if the accessed memory is not accessible, \ie the virtual address is invalid for the application, this raises an exception and terminate enclave execution. 
Hence, it is a challenge to obtain host address-space information stealthily from within an enclave.
To remain stealthy and avoid detection, the enclave needs a fault-resistant memory disclosure primitive.
Even with blind ROP~\cite{Bittau2014}, fault resistance may be necessary as pages are mapped on demand, and pagefaults would give away the ongoing attack.
\parhead{Achieving Fault Resistance}
For normal applications, fault resistance can be achieved by installing a user-space signal handler (on Linux) or structured exception handling (on Windows).
Upon an invalid memory access, the operating system delegates exception handling to the registered user-space handler.
Again, this is not possible from within an enclave due to limitations on syscalls.
Instead, we resemble this approach via TSX.

\subsection{TSX-based Address Probing}\label{sec:tmap}

We present a novel fault-resistant read primitive called \emph{TAP} (TSX-based Address Probing).\footnote{\label{ftn:code}The implementation can be found at \url{https://github.com/sgxrop/sgxrop}.}
In contrast to previous work, our attack is not a timing attack~\cite{Jang2016}, \ie we solely exploit the TSX error codes.
\emph{TAP} uses TSX to determine whether a virtual address is accessible by the current process (\ie mapped and user accessible) or not. 
\emph{TAP} exploits a side effect of TSX: When wrapping a memory access inside a TSX transaction, all potential access faults are caught by TSX instead of throwing an exception. 
Accessing an invalid memory location only aborts the transaction, but does not terminate the application. 
Thus, TSX allows to safely access any address within a transaction, without the risk of crashing the enclave. 
The resulting memory-disclosure primitive is extremely robust, as it automatically prevents reading of invalid memory locations. 
This has the advantage that an attacker does not require any knowledge of the memory layout, \ie which addresses are accessible.
\emph{TAP} probes an address as follows. 
We wrap a single read instruction to this address inside a TSX transaction.

\parhead{\emph{Accessible} Address}
If the accessed address is user-accessible, the transaction likely completes successfully.
In rare cases it might fail due to external influences, such as interrupts (\eg scheduling), cache eviction, or a concurrent modification of the accessed value. 
In these cases, TSX returns an error code indicating that the failure was only temporary and we can simply restart the transaction. 

\parhead{\emph{Inaccessible} Address}
If the address is inaccessible, TSX suppresses the exception~\cite{Intel_vol3} (\ie the operating system is not notified) and aborts the transaction. 
The user code receives an error code and can handle the transaction abort. 
Although the error code does not indicate the precise abort reason, it is distinct from temporary failures that suggest a retry.
Thus, we can deduce that the accessed address is either not mapped, or it is inaccessible from user space (\eg kernel memory).
Both reasons imply that the malicious enclave cannot read from the address.
Thus, a further distinction is not necessary. 

\parhead{TAP is Stealthy}
Although TSX can suppress exceptions from trapping to the operating system, TSX operation could be traced using hardware performance counters.
However, when running in enclave mode, most hardware performance counters are not updated~\cite{Schwarz2017SGX,Intel_SGXDifferences}.
We verified that especially none of the TSX-related performance counters are updated in enclave mode.
Thus, running TSX-based Address Probing~(\emph{TAP}) in enclave mode is entirely invisible to the operating system. 
Note that this primitive can also be used in regular exploits for ``egg hunting'', \ie scanning the address space for injected shellcode~\cite{Miller2004safely,Polychronakis2010shellcode}. 
As it does not rely on any syscalls, it can neither be detected nor prevented by simply blocking the syscalls typically used for egg hunting.

\subsection{Address-Space Exploration}\label{sec:as-template}

To mount a code-reuse attack, an attacker requires code gadgets to craft a chain of such gadgets. 
To collect enough gadgets, the enclave explores the host application's address space by means of \emph{TAP}.
Instead of applying \emph{TAP} to every probed address, it suffices to probe a single address per page.
This reveals whether the page is accessible to the enclave and allows the enclave to scan this entire page for gadgets via ordinary memory read instructions.

To detect gadgets, the attacker could in principle scan the entire virtual address space, which takes approximately 45 minutes (Intel i7-6700K).
To speed up the scanning, the attacker can apply JIT-ROP~\cite{Snow2013jitrop} to start scanning from a few known pointers.
For example, the malicious enclave knows the host code address to which the ECALL is supposed to return. 
Also, the stack pointer to the host application stack is visible to the enclave. 
By scanning the host stack, the enclave can infer further valid code locations, \eg due to saved return addresses.
Thus, \emph{TAP} can be used for the starting point of JIT-ROP, as well as to make JIT-ROP more resistant, as a wrongly inferred destination address does not crash the enclave. 

Although JIT-ROP is fast, the disadvantage is that it is complex and only finds a fraction of usable executable pages~\cite{Snow2013jitrop}. 
With \emph{TAP}, an attacker can choose the tradeoff between code coverage (\ie amount of discovered gadgets) and runtime of the gadget discovery. 
The most simple and complete approach approach is to linearly search through the entire virtual address space. 
To reduce the runtime of 45 minutes, an attacker can decide to use JIT-ROP for every mapped page instead of continuing to iterate through the address space.

After the address-space exploration, an attacker knows code pages which are usable to construct a ROP chain. 

\section{Escaping Enclaves with SGX-ROP}
\label{sec:rop} 
In this section, we present a novel way to mount a code-reuse attack from within SGX enclaves. 
We exploit the fact that SGX insufficiently isolates host applications from enclaves.
In particular, we show that the shared virtual address space between host application and enclave, in combination with our address-space exploration (\cf \Cref{sec:tsx}), allows an attacker to mount a code-reuse attack on the application. 
Subsequently, the attacker gains arbitrary code execution within the host application, even if it is well-written and bug-free. 

We discuss challenges in mounting the attack, and present solutions for all challenges. 
Moreover, we show how to construct a novel fault-resistant write primitive using TSX which allows an attacker to store additional shellcode. 

\subsection{Problem Statement}
The attacker wants to gain arbitrary code execution, which is typically achieved by loading attack code to a data page and then executing it. 
However, this requires syscalls to make the injected code executable. 
To mount the attack, the attacker first needs to escape the restricted enclave execution environment and bypass the host interface in order to execute syscalls.
Until now it was unclear whether and how this could be achieved in practice.
We show how to use SGX-ROP for that purpose.
To inject an SGX-ROP chain (or arbitrary code) into the host application, the attacker requires knowledge about which memory locations are writable. 
Similar to before (Section~\ref{sec:tsx}), this demands a fault-resistant method to detect writable memory pages.
Lastly, the attacker wants to remain stealthy and not perturb normal program execution.
In particular, the malicious enclave shall always perform benign operations it is supposed to do and shall return to its host application via the intended interface.
Also, after finishing the SGX-ROP attack, program execution shall continue normally. 

\subsection{Diverting the Control Flow}\label{sec:change-control-flow}

\parhead{Towards SGX-ROP}
In traditional code-reuse attacks, an attacker has to exploit a software bug (\eg a buffer overflow) to get control over the instruction pointer. 
However, due to the shared address space of the host application and the enclave, an attacker can access arbitrary host memory. 
Thus, the attacker implicitly has a \emph{write-anything-anywhere} primitive, allowing to directly overwrite instruction pointers, \eg stored on the stack. 
Since the attacker knows the precise location of the host stack, he can easily locate a saved instruction pointer on the host stack and prepare a code-reuse attack by replacing it with a ROP chain.
However, a code-reuse attack requires certain values to be on the current stack, \eg multiple return addresses and possibly function arguments.
Overwriting a larger part on the application stack might lead to data corruption and unexpected behavior of the host application. 
This would prevent recovering normal operation after the attack.
Moreover, in contrast to traditional control-flow hijacking attacks, an SGX-ROP attacker does not only want to manipulate the control flow but also completely restore the original control flow after the attack to preserve functionality and remain stealthy. 

Summing up, an SGX-ROP attacker cannot rely on any free or unused space on the current stack frame. 
Hence, the attacker requires a temporary stack frame to store the values required for the attack code. 

\parhead{Stealthy fake stack frames} 
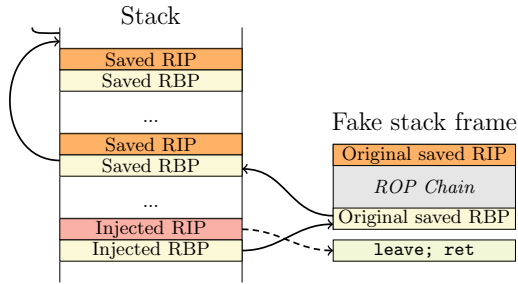
\begin{figure}[t]
 \centering
 \resizebox{0.6\hsize}{!}{
 \begin{tikzpicture}[yscale=.7]

\node at (1.5, 6.8) {\large Stack};

\draw (0,2) rectangle +(3, 2) node[pos=.5,yshift=-1.5em] {...};
\draw (0,4) rectangle +(3, 2) node[pos=.5,yshift=-1.5em] {...};
\draw (0,6) -- (0, 6.5);
\draw (3,6) -- (3, 6.5);
\draw (0,0.5) -- (0, 1.5);
\draw (3,0.5) -- (3, 1.5);

\draw[thick] (0,2) -- (3, 2);
\draw[thick] (0,4) -- (3, 4);
\draw[thick] (0,6) -- (3, 6);

\draw[fill=red!40] (0,1.5) rectangle +(3, 0.5) node[pos=.5] {Injected RIP};
\draw[fill=yellow!40] (0,1) rectangle +(3, 0.5) node[pos=.5] {Injected RBP};
\draw[fill=orange!60] (0,3.5) rectangle +(3, 0.5) node[pos=.5] {Saved RIP};
\draw[fill=yellow!40] (0,3) rectangle +(3, 0.5) node[pos=.5] {Saved RBP};
\draw[fill=orange!60] (0,5.5) rectangle +(3, 0.5) node[pos=.5] {Saved RIP};
\draw[fill=yellow!40] (0,5) rectangle +(3, 0.5) node[pos=.5] {Saved RBP};

\begin{scope}[yshift=1cm]
\node at (6, 3.3) {\large Fake stack frame};
\draw[fill=gray!20] (4.5,0.75) rectangle +(3, 2) node[pos=.5] {\textit{ROP Chain}};
\draw[fill=orange!60] (4.5,2.25) rectangle +(3, 0.5) node[pos=.5] {\small Original saved RIP};
\draw[fill=yellow!40] (4.5,0.75) rectangle +(3, 0.5) node[pos=.5] {\small Original saved RBP};

\draw[fill=green!20] (4.5,0) rectangle +(3, 0.5) node[pos=.5] {\texttt{leave; ret}};
\end{scope}

\draw[in=180,out=0,->,thick,densely dashed] (3, 1.75) to (4.5, 1.25);

\draw[in=180,out=0,->,thick] (3, 1.25) to (4.5, 1.87);
\draw[in=180,out=180,->,thick] (0, 3.37) to (0, 6.17);
\draw[in=260,out=180,thick] (0, 6.37) to (-0.45, 6.5);
\draw[in=0,out=180,->,thick] (4.5, 2.07) to (3, 3.17);

\end{tikzpicture}
 }
 \caption{To divert the control flow without interfering with legitimate stack frames, the attacker injects a new stack frame. The new stack frame can be used for arbitrary code-reuse attacks without leaving any traces in stack frames of other functions.}
 \label{fig:inject-frame}
 \vspace{-0.3cm}
\end{figure}
We present a technique to store the SGX-ROP chain on a temporary \emph{fake stack} which is an extension to stack pivoting~\cite{Prakash2015pivot}. 
The fake stack frame is located somewhere in unused writable memory, thus preserving stack data of the original program.
First, the attacker copies the saved instruction pointer and saved base pointer to the fake stack frame. 
Then, the attacker replaces the saved instruction pointer with the address of a function epilogue gadget, \ie \texttt{leave; ret}, and the saved base pointer with the address of the fake stack frame. 
With the pivot gadget, the stack is switched to the new fake stack frame. 
However, in contrast to a normal stack pivot, preservering the old values allows the attacker to resume with the normal control flow when returning from the fake stack. 
\Cref{fig:inject-frame} illustrates the stealthy stack pivoting process. 
The injected stack frame contains a ROP chain which is used as attack code and continues normal execution after the ROP chain was executed. 

If the compiler saves the base pointer on a function call, a fake stack frame can be placed between any two stack frames, not only after the current function.
Thus, attack code can be executed delayed, \ie not directly after the enclave returns to the host application, but at any later point where a function returns. 

SGX-ROP evades a variety of ROP defense mechanisms.
For example, stack canaries do not protect against SGX-ROP, since our fake stack frame bypasses the stack smashing detection.
For software-based shadow stacks without write protection~\cite{Chiueh2001,StackShield2011}, the attacker can perform SGX-ROP on the shadow stack as well.
The write-anything-anywhere primitive can also be leveraged to break CFI policies~\cite{Carlini2015control}, hardware-assisted CFI extensions~\cite{Theodorides2017breaking}, and stack-pivot defenses~\cite{Yan2016pivot}.

\parhead{Gaining arbitrary code execution}
With SGX-ROP, an attacker can stitch ROP gadgets together to execute syscalls in the host application.
To gain arbitrary code execution, the enclave can inject attacker payload on a writable page and then use the ROP chain to instruct the operating system to bypass execution prevention (\ie the non-executable bit). 
On Linux, this can be done with a single mprotect syscall.

\subsection{Detecting Writable Memory Locations}\label{sec:cave-mining}

For SGX-ROP, the attacker requires unused, writable host memory to inject a fake stack frame as well as arbitrary attack payload. 
The enclave cannot allocate host application memory for that purpose but instead attempts to misuse existing host memory.
However, as before, the attacker initially does not know the memory layout of its host application.
In this section, we present \emph{CLAW} (Checking Located Addresses for Writability), a combination of two TSX side effects to detect whether an arbitrary memory location is writable. 
This can be used to build a fault-resistant write primitive.

\begin{figure}[t]
 \centering
 \resizebox{0.8\hsize}{!}{
 \begin{tikzpicture}[xscale=1.2,yscale=1]
\draw[rounded corners,fill=gray!10] (-4,0.25) rectangle (4,-1.75);
\node at (3.25,0.05) {\texttt{TAP}};

\node (access) at (0,0) {\textbf{Read} \texttt{addr}};
\node[blue!60!black] (aabort) at (-3, -1.5) {Unmapped};
\node[blue!60!black] (aok) at (3, -1.5) {Mapped};
\node (cwrite) at (2.5, -2) {\textbf{Write} to \texttt{addr} + \texttt{xabort()}};

\draw[->,dashed,red!50!black] (access) -- (aabort) node [midway, above, sloped] {TSX fail};
\draw[->,thick,green!60!black] (access) -- (aok) node [midway, above, sloped] {TSX success};

\node[blue!60!black] (afail) at (-0.5, -3.5) {Read-only};
\node[blue!60!black] (aexplicit) at (5.5, -3.5) {Writable};

\draw[->,dashed,red!50!black] (cwrite) -- (afail) node [midway, above, sloped] {TSX fail};
\draw[->,thick,green!60!black] (cwrite) -- (aexplicit) node [midway, above, sloped] {TSX abort};

\end{tikzpicture}
 }
 \vspace{-0.3cm}
 \caption{\emph{CLAW} exploits that memory writes in TSX are only visible to the outside of a transaction if it succeeds, and that TSX distinguishes between implicit and explicit aborts. Thus, the return value of TSX after writing to an address and explicitly aborting determines whether the memory location is writable without changing it.}
 \label{fig:tsx-write}
 \vspace{-0.3cm}
\end{figure}
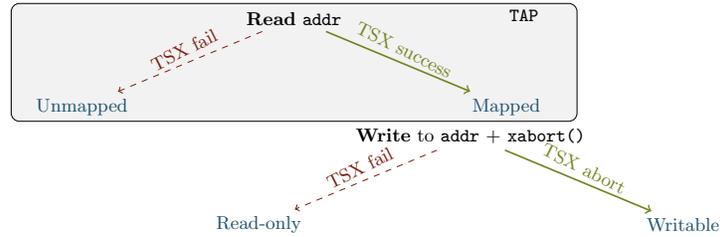

\emph{CLAW} first leverages \emph{TAP} to detect whether a virtual address is present, as shown in \Cref{fig:tsx-write}.\cref{ftn:code}
Then, \emph{CLAW} utilizes TSX to test whether this page is also writable.
To do so, we encapsulate a write instruction to the page of interest within a TSX transaction and explicitly abort the transaction after the write. 
Based on the return value of the transaction, we can deduce whether the page is writable. 
If the return value indicates an explicit abort, the write would have succeeded but was aborted by our explicit abort.
In this case, we can deduce that the page is writable. 
If the page is read-only, the transaction fails with an error distinct from an explicit abort.
The return value indicates that the transaction would never succeed, as the page is not writable. 
By observing those two error codes, one can distinguish read-only from writable pages, as shown in \Cref{fig:tsx-write}.

A property of \emph{CLAW} is that it is stealthy.
Since all memory writes within a transaction are only committed to memory if the transaction succeeds, our explicit abort ensures that memory remains unmodified.
Also, as with \emph{TAP}, \emph{CLAW} neither causes any exceptions to the operating system nor can it be seen in hardware performance counters.

\parhead{Fault-resistant write-anything-anywhere primitive}
With \emph{CLAW}, building a fault-resistant write primitive is straightforward.
Before writing to a page, \emph{CLAW} is leveraged to test whether the page is writable. 
Then, the content can be safely written to the page.

\parhead{Host Infection}
Both the fake stack frame as well as placing arbitrary attack payload (\eg a shellcode) require an unused writable memory location in the host application, which we denote as \emph{data cave}. 
After finishing address space exploration (\Cref{sec:as-template}), the malicious enclave uses \emph{CLAW} to test whether the found pages are writable. 
Again, probing a single address with \emph{CLAW} suffices to test whether the entire page is writable.
Moreover, the enclave needs to know whether it can safely use writable pages as data caves without destroying application data.
We consider empty pages (\ie they only contain `0's) as safe. 
Note that the ROP chain and possible shellcode should always be zeroed-out after execution to obscure the traces of the attack. 

\section{Attack Evaluation}\label{sec:evaluation} 
In this section, we evaluate \emph{TAP} and \emph{CLAW}, and show that \emph{TAP} can also be used in traditional exploits for egg hunting. 
We scan Graphene-SGX~\cite{Tsai2017Graphene} (an SGX wrapper library) for data caves and ROP gadgets and also scan the SGX SDK for ROP gadgets. 
Finally, we present a simple proof-of-concept exploit.
All evaluations were performed on an Intel i7-6700K with \SI{16}{\giga B} of memory. 

\subsection{\emph{TAP}+\emph{CLAW}}\label{sec:eval-tap} 

\begin{figure}[t]
 \centering
 \resizebox{\hsize}{!}{
 \begin{tikzpicture}[scale=1.2]
\newlength{\hatchspread}
\newlength{\hatchthickness}
\newlength{\hatchshift}
\newcommand{\hatchcolor}{}
\tikzset{hatchspread/.code={\setlength{\hatchspread}{#1}},
         hatchthickness/.code={\setlength{\hatchthickness}{#1}},
         hatchshift/.code={\setlength{\hatchshift}{#1}},
         hatchcolor/.code={\renewcommand{\hatchcolor}{#1}}}
\tikzset{hatchspread=3pt,
         hatchthickness=0.4pt,
         hatchshift=0pt,
         hatchcolor=black}
\pgfdeclarepatternformonly[\hatchspread,\hatchthickness,\hatchshift,\hatchcolor]
   {custom north west lines}
   {\pgfqpoint{\dimexpr-2\hatchthickness}{\dimexpr-2\hatchthickness}}
   {\pgfqpoint{\dimexpr\hatchspread+2\hatchthickness}{\dimexpr\hatchspread+2\hatchthickness}}
   {\pgfqpoint{\dimexpr\hatchspread}{\dimexpr\hatchspread}}
   {
    \pgfsetlinewidth{\hatchthickness}
    \pgfpathmoveto{\pgfqpoint{0pt}{\dimexpr\hatchspread+\hatchshift}}
    \pgfpathlineto{\pgfqpoint{\dimexpr\hatchspread+0.15pt+\hatchshift}{-0.15pt}}
    \ifdim \hatchshift > 0pt
      \pgfpathmoveto{\pgfqpoint{0pt}{\hatchshift}}
      \pgfpathlineto{\pgfqpoint{\dimexpr0.15pt+\hatchshift}{-0.15pt}}
    \fi
    \pgfsetstrokecolor{\hatchcolor}
    \pgfusepath{stroke}
   }

\draw[fill=gray!40] (0.00,0) rectangle +(1.50,10);
\node[rotate=45,xshift=-3.8cm,yshift=-2.8cm,anchor=west,scale=3] at (0, 14) {\texttt{400000}};
\node[rotate=90,scale=3] at (0.75, 5) {Binary};
\draw[pattern=custom north west lines,hatchcolor=gray!40,hatchspread=20pt,hatchthickness=2pt] (1.50,5) rectangle +(15.72,5);
\node[rotate=90,scale=3] at (9.36, 7.5) {Binary};
\draw[fill=gray!40] (17.22,0) rectangle +(1.50,10);
\node[rotate=90,scale=3] at (17.97, 5) {Binary};
\draw[decorate,thick,decoration={snake,amplitude=10,segment length=120,post length=0},draw=black!80!white] (19.22,0) -- (19.22,10);
\node[rotate=45,xshift=-3.8cm,yshift=-2.8cm,anchor=west,scale=3] at (18.72, 14) {\texttt{802000}};
\draw[decorate,thick,decoration={snake,amplitude=10,segment length=120,post length=0},draw=black!80!white] (20.72,0) -- (20.72,10);
\node[rotate=45,xshift=-3.8cm,yshift=-2.8cm,anchor=west,scale=3] at (21.22, 14) {\texttt{a01000}};
\draw[fill=gray!40] (21.22,0) rectangle +(1.50,10);
\node[rotate=90,scale=3] at (21.97, 5) {Binary};
\draw[decorate,thick,decoration={snake,amplitude=10,segment length=120,post length=0},draw=black!80!white] (23.22,0) -- (23.22,10);
\node[rotate=45,xshift=-3.8cm,yshift=-2.8cm,anchor=west,scale=3] at (22.72, 14) {\texttt{a03000}};
\draw[decorate,thick,decoration={snake,amplitude=10,segment length=120,post length=0},draw=black!80!white] (24.72,0) -- (24.72,10);
\node[rotate=45,xshift=-3.8cm,yshift=-2.8cm,anchor=west,scale=3] at (25.22, 14) {\texttt{bfe000}};
\draw[fill=yellow!40] (25.22,0) rectangle +(1.50,10);
\node[rotate=90,scale=3] at (25.97, 5) {Heap};
\draw[fill=yellow!20] (26.72,5) rectangle +(1.50,5);
\draw[pattern=custom north west lines,hatchcolor=yellow!80!black,hatchspread=20pt,hatchthickness=2pt] (26.72,5) rectangle +(1.50,5);
\node[rotate=90,scale=3] at (27.47, 7.5) {Heap};
\draw[decorate,thick,decoration={snake,amplitude=10,segment length=120,post length=0},draw=black!80!white] (28.72,0) -- (28.72,10);
\node[rotate=45,xshift=-3.8cm,yshift=-2.8cm,anchor=west,scale=3] at (28.22, 14) {\texttt{c1f000}};
\draw[decorate,thick,decoration={snake,amplitude=10,segment length=120,post length=0},draw=black!80!white] (30.22,0) -- (30.22,10);
\node[rotate=45,xshift=-3.8cm,yshift=-2.8cm,anchor=west,scale=3] at (30.72, 14) {\texttt{7fded9327000}};
\draw[fill=gray!40] (30.72,0) rectangle +(1.50,10);
\node[rotate=90,scale=3] at (31.47, 5) {libc};
\draw[pattern=custom north west lines,hatchcolor=gray!40,hatchspread=20pt,hatchthickness=2pt] (32.22,5) rectangle +(1.50,5);
\node[rotate=90,scale=3] at (32.97, 5) {};
\draw[fill=gray!40] (33.72,0) rectangle +(1.50,10);
\node[rotate=90,scale=3] at (34.47, 5) {};
\draw[pattern=custom north west lines,hatchcolor=gray!40,hatchspread=20pt,hatchthickness=2pt] (35.22,5) rectangle +(1.50,5);
\node[rotate=90,scale=3] at (35.97, 5) {};
\draw[fill=gray!40] (36.72,0) rectangle +(1.50,10);
\node[rotate=90,scale=3] at (37.47, 5) {};
\draw[pattern=custom north west lines,hatchcolor=gray!40,hatchspread=20pt,hatchthickness=2pt] (38.22,5) rectangle +(1.50,5);
\node[rotate=90,scale=3] at (38.97, 5) {};
\draw[fill=gray!40] (39.72,0) rectangle +(1.50,10);
\node[rotate=90,scale=3] at (40.47, 5) {};
\draw[pattern=custom north west lines,hatchcolor=gray!40,hatchspread=20pt,hatchthickness=2pt] (41.22,5) rectangle +(1.50,5);
\node[rotate=90,scale=3] at (41.97, 5) {};
\draw[fill=gray!40] (42.72,0) rectangle +(1.50,10);
\node[rotate=90,scale=3] at (43.47, 5) {};
\draw[pattern=custom north west lines,hatchcolor=gray!40,hatchspread=20pt,hatchthickness=2pt] (44.22,5) rectangle +(1.50,5);
\node[rotate=90,scale=3] at (44.97, 5) {};
\draw[fill=gray!40] (45.72,0) rectangle +(1.50,10);
\node[rotate=90,scale=3] at (46.47, 5) {ldd};
\draw[pattern=custom north west lines,hatchcolor=gray!40,hatchspread=20pt,hatchthickness=2pt] (47.22,5) rectangle +(8.58,5);
\node[rotate=90,scale=3] at (51.51, 7.5) {ldd};
\draw[fill=gray!40] (55.80,0) rectangle +(1.50,10);
\node[rotate=90,scale=3] at (56.55, 5) {ldd};
\draw[fill=gray!40] (57.30,0) rectangle +(1.50,5);
\node[rotate=90,scale=3] at (58.05, 5) {};
\draw[decorate,thick,decoration={snake,amplitude=10,segment length=120,post length=0},draw=black!80!white] (59.30,0) -- (59.30,10);
\node[rotate=45,xshift=-3.8cm,yshift=-2.8cm,anchor=west,scale=3] at (58.80, 14) {\texttt{7fded9718000}};
\draw[decorate,thick,decoration={snake,amplitude=10,segment length=120,post length=0},draw=black!80!white] (60.80,0) -- (60.80,10);
\node[rotate=45,xshift=-3.8cm,yshift=-2.8cm,anchor=west,scale=3] at (61.30, 14) {\texttt{7fded98e5000}};
\draw[fill=yellow!40] (61.30,0) rectangle +(1.50,10);
\node[rotate=90,scale=3] at (62.05, 5) {unknown};
\draw[fill=yellow!40] (63.48,0) rectangle +(1.50,10);
\node[rotate=90,scale=3] at (64.23, 5) {ldd};
\draw[decorate,thick,decoration={snake,amplitude=10,segment length=120,post length=0},draw=black!80!white] (65.48,0) -- (65.48,10);
\node[rotate=45,xshift=-3.8cm,yshift=-2.8cm,anchor=west,scale=3] at (64.98, 14) {\texttt{7fded9919000}};
\draw[decorate,thick,decoration={snake,amplitude=10,segment length=120,post length=0},draw=black!80!white] (66.98,0) -- (66.98,10);
\node[rotate=45,xshift=-3.8cm,yshift=-2.8cm,anchor=west,scale=3] at (67.48, 14) {\texttt{7ffcf0fc7000}};
\draw[fill=yellow!20] (67.48,5) rectangle +(1.50,5);
\draw[pattern=custom north west lines,hatchcolor=yellow!80!black,hatchspread=20pt,hatchthickness=2pt] (67.48,5) rectangle +(1.50,5);
\node[rotate=90,scale=3] at (68.23, 7.5) {Stack};
\draw[fill=yellow!40] (68.98,0) rectangle +(1.50,10);
\node[rotate=90,scale=3] at (69.73, 5) {Stack};
\draw[fill=gray!40] (70.48,0) rectangle +(1.50,5);
\node[rotate=90,scale=3] at (71.23, 2.5) {VVar};
\draw[fill=gray!40] (71.98,0) rectangle +(1.50,10);
\node[rotate=90,scale=3] at (72.73, 5) {vDSO};
\draw[decorate,thick,decoration={snake,amplitude=10,segment length=120,post length=0},draw=black!80!white] (73.98,0) -- (73.98,10);
\node[rotate=45,xshift=-3.8cm,yshift=-2.8cm,anchor=west,scale=3] at (73.48, 14) {\texttt{7ffcf0fee000}};
\draw[decorate,thick,decoration={snake,amplitude=10,segment length=120,post length=0},draw=black!80!white] (75.48,0) -- (75.48,10);
\node[rotate=45,xshift=-3.8cm,yshift=-2.8cm,anchor=west,scale=3] at (75.98, 14) {\texttt{f'ffff600000}};
\draw[fill=gray!40] (75.98,0) rectangle +(1.50,10);
\node[rotate=90,scale=3] at (76.73, 5) {VSyscall};
\draw[decorate,thick,decoration={snake,amplitude=10,segment length=120,post length=0},draw=black!80!white] (77.98,0) -- (77.98,10);
\node[rotate=45,xshift=-3.8cm,yshift=-2.8cm,anchor=west,scale=3] at (77.48, 14) {\texttt{f'ffff601000}};
\draw[decorate,thick,decoration={snake,amplitude=10,segment length=120,post length=0},draw=black!80!white] (79.48,0) -- (79.48,10);
\node[rotate=45,xshift=-3.8cm,yshift=-2.8cm,anchor=west,scale=3] at (79.98, 14) {\texttt{f'ffffffffff}};
\draw[fill=none,thick] (0,0) rectangle (79.98,10);
\draw[ultra thick,dashed,red] (-2,5) -- (82,5);
\node[rotate=90,scale=3] at (-1,7.5) {Linux};
\node[rotate=90,scale=3] at (-1,2.5) {TAP+CLAW};
\end{tikzpicture}
 }
 \caption{The virtual memory layout of a simple program on Linux (x86\_64) as provided by \texttt{/proc/<pid>/maps} (top) and reconstructed using \emph{TAP}+\emph{CLAW} (bottom).}
 \label{fig:tsx-compare}
\vspace{-0.3cm}
\end{figure}
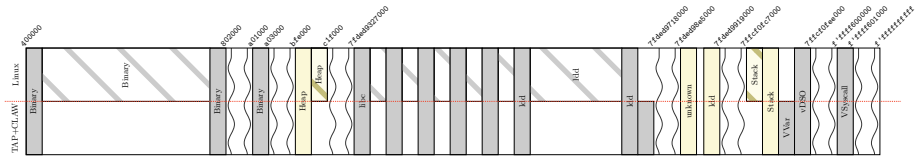

We used the combination \emph{TAP}+\emph{CLAW} to scan the virtual memory of a process and also distinguish writable from read-only pages. 
\Cref{fig:tsx-compare} shows the memory map of a process recovered with \emph{TAP}+\emph{CLAW} (bottom), compared with the ground truth directly obtained from the procfs file system (top). 
The procfs file system shows more areas (shaded), as it also includes pages which are not mapped in memory, but only defined in the binary. 
All mapped pages were correctly identified using \emph{TAP}, and also the distinction between read-only and writable pages using \emph{CLAW} was always correct. 

Both \emph{TAP} and \emph{CLAW} are very fast, taking only the time of a cache read or write (around \SIx{330} cycles for an uncached memory access on our test machine) plus the overhead for a TSX transaction, which we measured as \SIx{30} cycles. 
Scanning the entire virtual address space takes \SI{45}{\minute}, resulting in a probing rate of \SI{48.5}{\giga\byte/\second}. 
To estimate the runtime of \emph{TAP} and \emph{CLAW} on real-world applications, we evaluted both primitives on the 97 GNU Core Utilities with ASLR enabled. 
We linearly explored the binary starting from one known address (similarly to JIT-ROP~\cite{Snow2013jitrop}). 
On average, \emph{TAP} located all pages of the application within \SI{73.5}{\milli\second}. 
This time can be reduced further, as an attacker can stop probing as soon as all required gadgets are found. 

\textbf{Egg Hunting.}
We also evaluated \emph{TAP} as a novel method for egg hunting in regular (non-enclave) attacks, \ie scanning the address space for injected shellcode~\cite{Miller2004safely,Polychronakis2010shellcode}. 
State-of-the-art egg hunters for Linux~\cite{Nemeth2015egghunt,Miller2004safely} rely on syscalls (\eg \texttt{access}) which report whether one of the parameters is a valid address. 
However, issuing a syscall requires the \texttt{syscall} instruction as well as setting up the required parameters. 
Thus, such egg hunters are usually larger than \SIx{30} bytes~\cite{Miller2004safely}. 
Nemeth~\etal\cite{Nemeth2015egghunt} argued that egg hunters with fault prevention cannot be smaller.
However, our \emph{TAP} egg hunter is only \SIx{16} bytes in size,\cref{ftn:code} \ie the smallest egg hunter with fault prevention.
With \SIx{360} cycles per address, it is also significantly faster (by a factor of \SIx{4.8}) than egg hunters leveraging the \texttt{access} syscall (\SIx{1730} cycles per address).

\subsection{Code-reuse Gadgets and Data Caves in SGX Frameworks}

To evaluate the viability of a code-reuse attack using a fake stack frame (\cf \Cref{sec:change-control-flow}), we inspected Graphene-SGX for data caves (\cf \Cref{sec:cave-mining}) and ROP gadgets. 
We chose Graphene-SGX, as it is open source\footnote{\url{https://github.com/oscarlab/graphene}}, actively maintained, and designed to run unmodified binaries within SGX~\cite{Tsai2017Graphene}. 
Furthermore, we also analyzed the Intel SGX SDK for ROP gadgets, as it is the most common framework for SGX applications. 

Our simple attack enclave used \emph{TAP}+\emph{CLAW} to find code pages and data caves. 
We successfully detected all mapped pages of the host application, and also distinguished between writable and read-only pages. 

\parhead{Data Caves}
With \emph{CLAW}, we were able to detect which pages are not only present but also writable. 
For the writable pages, we further analyzed whether they contain only `0's and are thus data caves. 
We found \SIx{16594} data caves in Graphene-SGX, which took on average \SI{45.5}{\milli\second}. 
This amounts to around \SI{64.8}{\mega B} of memory which can be used by an attacker. 
Such data caves also exist in the Intel SGX SDK. 
Thus, even highly complex malware such as zero-day exploits can be stored there. 
For traditional shellcode, a one-page data cave is already sufficient, as such a shellcode fits within a few bytes. 

\parhead{Gadgets}
Data caves allow storing arbitrary code to the memory, and an attacker only has to mark them as executable to run the stored code. 
Thus, to make use of the data caves, an attacker requires a ROP chain which makes the data caves executable, \eg the \texttt{mprotect} syscall on Linux. 
This syscall can be called using a ROP chain consisting of only 4 gadgets: \texttt{POP RDI}, \texttt{POP RSI}, \texttt{POP RAX}, and \texttt{SYSCALL}. 
While these gadgets could also be replaced by more complex ones~\cite{Shacham2007}, they suffice for our purpose. 
We analyzed the mapped code pages of Graphene-SGX which we identified using \emph{TAP} (\cf \Cref{sec:as-template}). 
We found complete \texttt{mprotect} ROP chains (\ie all 4 gadgets) in multiple mapped code pages of Graphene-SGX. 
These pages include the binary itself (pal-linux), the math library (libm), the GNU C library (libc) and the GNU linker (ld). 

Furthermore, gadgets for ROP chains can be in different libraries.
For example, 3 out of the 4 gadgets are not only in one of the core libraries of Graphene-SGX, namely libsysdb, but also in the Intel SGX SDK itself (libsgx\_urts). 
The fourth gadget (\texttt{SYSCALL}) to build a complete chain can, \eg be found in the virtual syscall page, which is mapped in every process on modern Linux systems, or in the libc. 
Since any SGX application host is linked against libsgx\_urts and libc, building an \texttt{mprotect} ROP chain is always possible.

\subsection{Full Exploit}\label{sec:full-exploit}
Our proof-of-concept exploit consists of a benign application hosting a malicious enclave.
We use the most restricted enclave interface possible: the enclave may not use any OCALLs. 
After entering the enclave via any ECALL, the enclave uses \emph{TAP} and \emph{CLAW} to find and inject code and data into a data cave.
Using \emph{TAP}, the enclave detects host binary pages and automatically builds a ROP chain which creates a new file (in a real attack, the enclave would encrypt existing files) and displays a ransom message. 
We divert the control flow (\cf \Cref{sec:change-control-flow}) to let the host application execute the ROP chain, clean up stacks and immediately continue normal execution. 
The data cave provides enough space for sophisticated exploits and post-exploitation code, \eg complete zero-day exploits, or remote shells which can be used for further exploitation.

Our host application uses ASLR, stack canaries, and address sanitizer. 
The host application does not provide any addresses to the enclave which can be used as a starting point. 
Still, the end-to-end exploit\cref{ftn:code} takes on average only \SI{20.8}{\second}. 

\section{Discussion}\label{sec:discussion} 

SGX-ROP surpasses traditional ROP attacks, as the enclave isolation works only in one direction, \ie the enclave is protected from the host application, but not vice-versa. 
A write-anything-anywhere primitive is sufficient to break even extremely strict CFI policies~\cite{Carlini2015control} and hardware-assisted control-flow integrity extensions~\cite{Theodorides2017breaking}. 
In contrast to regular ROP attacks, we do not require a memory safety violation.
Also, the Intel SGX SDK yields enough ROP gadgets and data caves to gain arbitrary code execution.
Hence, SGX-ROP is always possible on current applications if, inadvertently, a malicious enclave is embedded.

With SGX-ROP, porting malware to SGX becomes trivial, thus intensifying the threat of enclave malware.
Moreover, hiding malware in an SGX enclave give attackers plausible deniability and stealthiness until they choose to launch the attack.
This is particularly relevant for trigger-based malware that embeds a zero-day exploit, but also to provide plausible deniability for legal or political reasons, \eg for a state actor~\cite{EFF2011NSAtrojan,Bundestrojaner2017Germany}. 
Possible scenarios range from synchronized large-scale denial-of-service attacks to targeted attacks on individuals.

\section{Conclusion}\label{sec:conclusion} 
We practically demonstrated the first enclave malware which fully and stealthily impersonates its host application.
Our attack uses new TSX-based techniques: a memory-disclosure primitive and a write-anything-anywhere primitive.
With SGX-ROP, we bypassed ASLR, stack canaries, and address sanitizer, to run ROP gadgets in the host context enabling practical enclave malware.
We conclude that instead of protecting users from harm, SGX currently poses a security threat, facilitating so-called super-malware with ready-to-hit exploits. 
Our results lay ground for future research on more realistic trust relationships between enclave and non-enclave software, as well as the mitigation of enclave malware.

\bibliographystyle{splncs04}
\bibliography{main}

\end{document}